\documentclass[conference,romanappendices,10pt]{IEEEtran}
\IEEEoverridecommandlockouts

\usepackage{algorithm,algorithmic,amsmath,amssymb,amsthm,array,bbm,bibentry,cite,color,comment}
\usepackage{enumerate,enumitem,eurosym,float,graphicx,lettrine,mathrsfs,mathtools,multirow,nomencl}
\usepackage{pict2e,psfrag,ragged2e,setspace,subfigure,supertabular,tabularx,url}

\usepackage[english]{babel}

{		\newtheorem{theorem}{Theorem}}
{ 		}
{ 		}
{ 		}
{ 		}
{		}
{		}
{ 		\newtheorem{corollary}{Corollary}}

\newcommand{\h}{\mathbf{h}}

\renewcommand{\v}{\mathbf{v}}

\renewcommand{\H}{\mathbf{H}}
\newcommand{\I}{\mathbf{I}}

\renewcommand{\L}{\mathbf{L}}

\newcommand{\setC}{\mathcal{C}}

\newcommand{\setL}{\mathcal{L}}

\newcommand{\setN}{\mathcal{N}}

\newcommand{\Real}{\mbox{$\mathbb{R}$}}
\newcommand{\Compl}{\mbox{$\mathbb{C}$}}

\newcommand{\argmin}{\operatornamewithlimits{argmin}}

\newcommand{\diff}{\mathrm{d}}
\newcommand{\Exp}{\mathbb{E}}
\newcommand{\herm}{\mathrm{H}}
\renewcommand{\Pr}{\mathbb{P}}

\newcommand{\SIR}{\mathsf{SIR}}
\newcommand{\suc}{\mathrm{suc}}

\DeclarePairedDelimiter{\ceil}{\lceil}{\rceil}

\title{Performance Analysis of Partial Interference Cancellation in Multi-Antenna UDNs}
\author{\IEEEauthorblockN{Italo Atzeni and Marios Kountouris}
\IEEEauthorblockA{Mathematical and Algorithmic Sciences Lab \\ France Research Center, Huawei Technologies Co., Ltd. \\
Email: \{italo.atzeni, marios.kountouris\}@huawei.com}}

\begin{document}

\maketitle

\begin{abstract}
The employment of partial zero-forcing (PZF) receivers at the base stations represents an efficient and low-complexity technique for uplink interference management in cellular networks. In this paper, we focus on the performance analysis of ultra-dense networks (UDNs) in which the multi-antenna receivers adopt PZF. We provide both integral expressions and tight closed-form approximations for the probability of successful transmission, which can be used to accurately evaluate the optimal tradeoff between interference cancellation and array gain. Numerical results show that no more than half of the available degrees of freedom should be used for interference cancellation.
\end{abstract}



\section{Introduction} \label{sec:Intro}


Ultra-dense networks (UDNs), i.e., dense and massive deployment of small cells that have a wired/wireless backhaul connection, represent a core element of emerging 5th generation (5G) wireless systems, which is expected to cope with the proliferation of wireless devices and the ever-growing demand for high data rates. Due to such massive and dense deployments, modern cellular networks are becoming interference-limited, thus motivating the use of interference management techniques \cite{Hua12, Hos15} such as successive interference cancellation (SIC) at both base stations (BSs) and user terminals. SIC receivers attempt to decode and subtract/cancel interfering signals in the order of decreasing interference power level and are characterized by a considerable complexity (see, e.g., \cite{Zha13b,Zha14b}) compared to linear processing.

In presence of multiple antennas at the receive side, the partial zero-forcing (PZF) receiver represents an efficient and low-complexity alternative to SIC for interference management \cite{Jin08,Ako11}. If a node is equipped with $N_{\mathrm{R}}$ receive antennas, PZF cancels the interference coming from $M \leq N_{\mathrm{R}} - 1$ interferers while using the remaining degrees of freedom to boost the desired received signal. Despite its evident advantage in terms of complexity, studying the performance of PZF receivers in UDNs using tools from stochastic geometry proves to be more troublesome than SIC since the presence of multiple antennas leads to generally complicated and often intractable expressions (see, for instance, \cite{Atz15a}).

In this paper, we present results on the performance analysis of dense random wireless networks, where the multi-antenna receivers adopt PZF for uplink interference management. More specifically, using a stochastic geometry-based framework, we provide both integral expressions and tight closed-form approximations for the probability of successful transmission (also termed as success probability). These tractable expressions can be used to accurately evaluate the optimal balance/tradeoff between interference cancellation and array gain. Numerical results show that no more than half of the available degrees of freedom should be used for interference cancellation.

\section{System Model} \label{sec:SM}
\subsection{Network Model} \label{sec:SM_NS}

Consider a reference receiver located at the origin of the Euclidean plane and its associated transmitter located at $x_{0} \in \Real^{2}$, and assume a fixed distance $R_{0} \triangleq \|x_{0}\|$ between the two nodes. The reference node receives interference coming from a set of transmitting nodes, whose location distribution is modeled according to the stationary Poisson point process (PPP) $\Phi \triangleq \{ x_{i} \}_{i=1}^{\infty} \subset \Real^{2}$ with spatial density $\lambda$ (measured in [nodes/m$^2$]). Let $X_{i} \triangleq \|x_{i}\|$, $\forall x_{i} \in \Phi$: without loss of generality, we assume that the points of $\Phi$ are indexed such that their distances from the reference node are in increasing order, i.e., $\{X_{i} \leq X_{i+1}\}_{i=1}^{\infty}$. Thanks to Slivnyak's theorem \cite[Ch.~8.5]{Hae12} and to the stationarity of $\Phi$, the statistics of the signal reception at the reference receiver (typical node) are the same for any receiver in the network.

\subsection{Channel Model} \label{sec:SM_CM}

In our setting, the reference receiver is equipped with $N_{\mathrm{R}}$ receive antennas, whereas its associated transmitter and the interfering nodes have a single transmit antenna.\footnote{This scenario can also model the case where transmitting nodes with multiple transmit antennas send a single stream using transmit beamforming techniques.} Furthermore, we assume that the associated transmitter $x_{0}$ has transmit power $\rho_{0}$ and that all the interfering nodes $\Phi$ transmit with the same power $\rho$. The propagation through the wireless channel is characterized as the combination of pathloss attenuation and small-scale fading. For the former, we consider the standard power-law pathloss model given by the function $\ell(x_{j}) \triangleq X_{j}^{-\alpha}$ with pathloss exponent $\alpha > 2$. For the latter, we assume Rayleigh fading and use $\h_{j} \sim \setC \setN (0, \I)$ to denote the channels from node $x_{j}$ to the reference receiver, where $\I$ is the $N_{\mathrm{R}}$-dimensional identity matrix.

\subsection{Partial Zero-Forcing Receiver} \label{sec:SM_PZF}

Assume now that the reference receiver employs a PZF receiver \cite{Jin08} to cancel the $M$ nearest interfering nodes: this approach is relevant when the order statistics of the received signal power are dominated by the pathloss rather than by the small-scale fading \cite{Wil14}. In this setting, let us define $\Compl^{N_{\mathrm{R}} \times M} \ni \H(M) \triangleq (\h_{1} \; \ldots \; \h_{M})$, which includes the effective channels from the $M$ nearest interfering nodes, and let us introduce the receive combining vector $\v(M) \in \Compl^{N_{\mathrm{R}}}$ defined as
\begin{align}
\label{eq:v} \v(M) \triangleq \frac{(\I - \H(M) \H^{\sharp}(M)) \h_{0}}{\| (\I - \H(M) \H^{\sharp}(M)) \h_{0} \|}.
\end{align}
Observe that $\v(M)$ in \eqref{eq:v} reduces to maximum ratio combining (MRC) when $M = 0$ and to full zero forcing when $M = N_{\mathrm{R}} - 1$.

Considering the network to be interference-limited (background noise is ignored), the signal-to-interference ratio (SIR) at the reference receiver when PZF is adopted is given by
\begin{align}
\label{eq:SIR} \SIR(M) \triangleq \frac{\rho_{0} R_{0}^{-\alpha} S_{0}(M)}{I(M)}
\end{align}
where we have defined $S_{j}(M) \triangleq |\v^{\herm}(M) \h_{j}|^{2}$ and
\begin{align}
\label{eq:I} I(M) \triangleq \rho \sum_{i = M+1}^{\infty} X_{i}^{-\alpha} S_{i}(M)
\end{align}
denotes the overall interference term.

\section{Probability of Successful Transmission} \label{sec:SP}
\subsection{Partial Interference Cancellation} \label{sec:SP_IC}

In this section, we provide the expression for the success probability with partial interference cancellation. The successful decoding by the reference receiver of a packet coming from its associated transmitter $x_{0}$ is defined as $\mathsf{P}_{\suc}(\theta,M) \triangleq \Pr[\SIR(M) > \theta]$ for a given SIR threshold $\theta$. Observe that $\mathsf{P}_{\suc}(\theta,M)$ corresponds to the the complementary cumulative distribution function (CCDF) of $\SIR$. The success probability when PZF is adopted by the reference receiver is formalized in the following theorem.

\begin{theorem} \label{th:P_suc1} \rm{
Assume that the reference receiver uses PZF to cancel the nearest $M$ interfering nodes. Then, the success probability is given by
\begin{align} \label{eq:P_suc1}
\mathsf{P}_{\suc}(\theta,M) = \! \sum_{n=0}^{N_{\mathrm{R}} - M - 1} \bigg[ \frac{(- s)^{n}}{n!} \frac{\mathrm{d}^{n}}{\mathrm{d} s^{n}} \setL_{I(M)}(s) \bigg]_{s = \theta \rho_{0}^{-1} R_{0}^{\alpha}}
\end{align}
where
\begin{align} \label{eq:LI}
\setL_{I(M)}(s) \triangleq \Exp_{\Phi} \Bigg[ \prod_{i = M+1}^{\infty} \frac{1}{1+s \rho X_{i}^{-\alpha}} \Bigg]
\end{align}
is the Laplace transform of the interference $I(M)$ in \eqref{eq:I}. For $M=0$, the Laplace transform has the closed-form expression
\begin{align} \label{eq:LI_0}
\setL_{I(0)}(s) = \exp \bigg( -2 \pi^{2} \lambda \frac{(s \rho)^{\frac{2}{\alpha}}}{\alpha} \csc \bigg( \frac{2 \pi}{\alpha} \bigg) \bigg).
\end{align}}
\end{theorem}

\begin{IEEEproof}
See Appendix~\ref{sec:A_P_suc1_th}.
\end{IEEEproof} \vspace{1mm}

\noindent Theorem~\ref{th:P_suc1} highlights the tradeoff between interference cancellation and array gain: the larger is $M$, the larger is $\setL_{I(M)}(s)$ in \eqref{eq:LI} (i.e., the lower is the interference), but also the less terms are included in the summation of $\mathsf{P}_{\suc}(\theta,M)$ in \eqref{eq:P_suc1} (note that all such terms are positive since the $n$-derivatives of $\setL_{I(M)}(s)$ are negative for odd $n$).

\begin{figure*}[t!]
\centering
\includegraphics[scale=1]{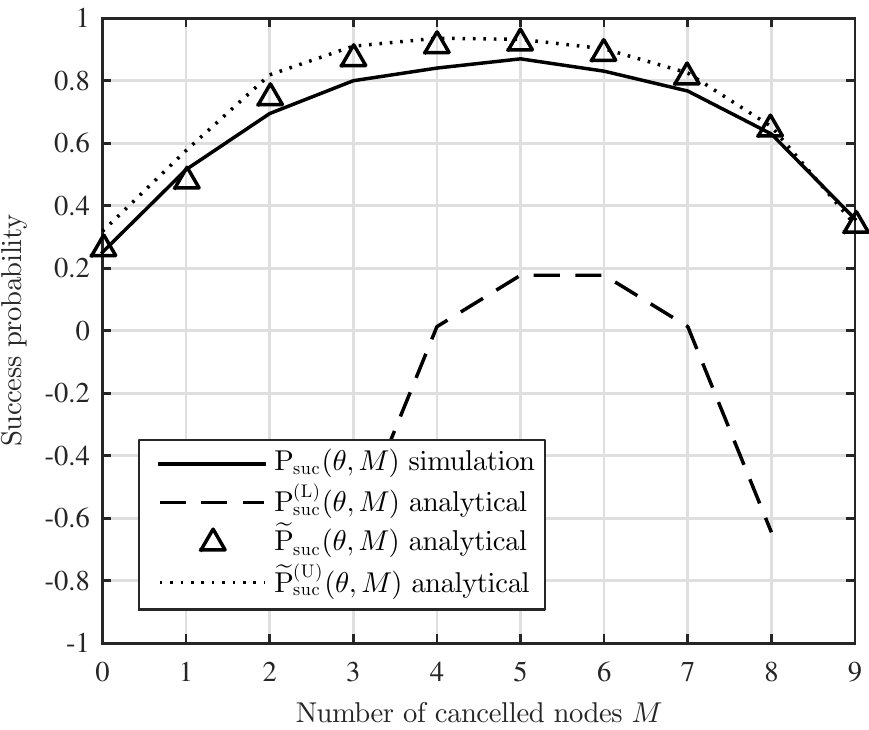} \hspace{2mm}
\includegraphics[scale=1]{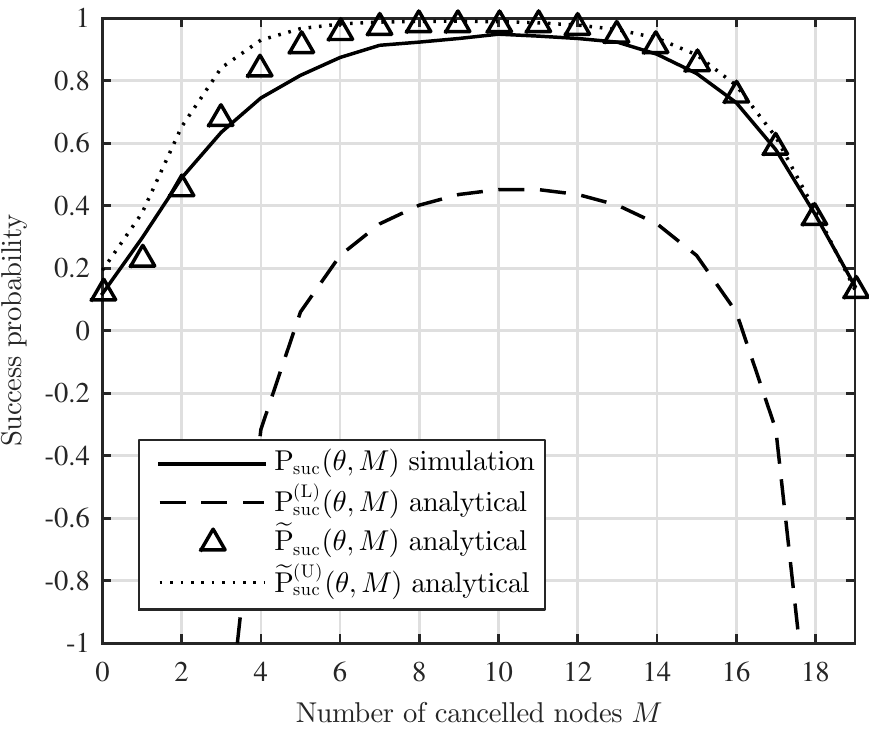} \vspace{-2mm}
\caption{Success probability against the number of cancelled interfering nodes $M$ with $\theta = 1$: $\lambda = 10^{-2}$~nodes/m$^2$ and $N_{\mathrm{R}}=10$ (left), $\lambda = 2 \times 10^{-2}$~nodes/m$^2$ and $N_{\mathrm{R}}=20$ (right).} \label{fig:comp1} \vspace{-1mm}
\end{figure*}

\begin{figure}[t!]
\centering
\includegraphics[scale=1]{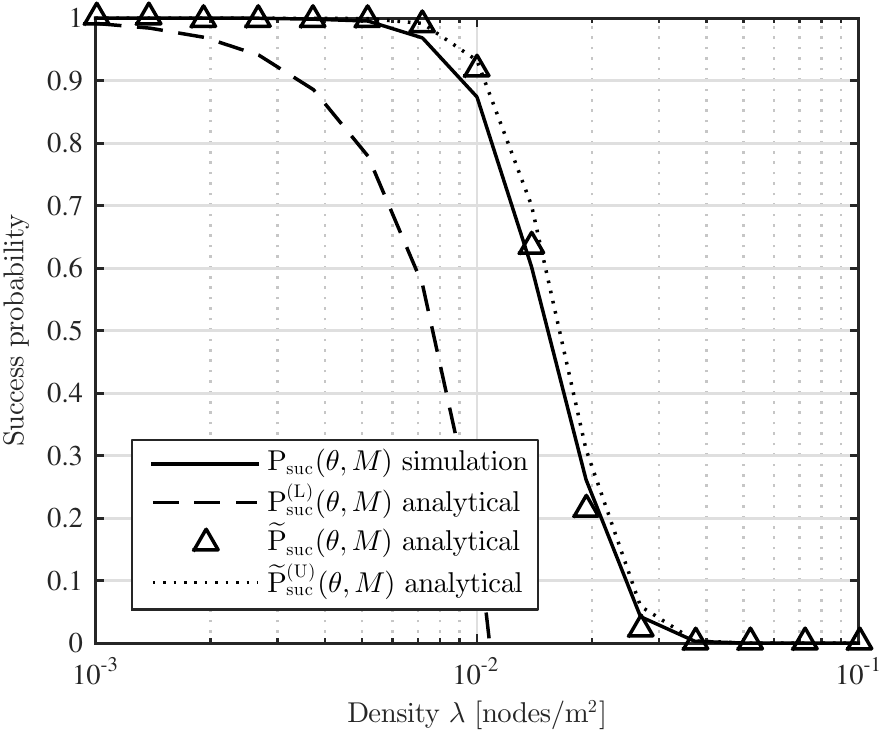} \vspace{-2mm}
\caption{Success probability versus density $\lambda$ with $N_{\mathrm{R}}=10$ and $M=5$.} \label{fig:comp2}
\end{figure}

Unfortunately, $\mathsf{P}_{\suc}(\theta,M)$ provided in Theorem~\ref{th:P_suc1} is not given in closed form since the expression of the Laplace transform $\setL_{I(M)}(s)$ is not available for $M \geq 1$. Resorting to \cite{Jin08}, one can readily obtain the following tractable lower bound for $M \in \big[ \lceil \tfrac{\alpha}{2} \rceil + 1, N_{\mathrm{R}} -2 \big]$: we have that $\mathsf{P}_{\suc}(\theta,M) > \mathsf{P}_{\suc}^{(\mathrm{L})}(\theta,M)$ with
\begin{align} \label{eq:P_suc1_LB}
\mathsf{P}_{\suc}^{(\mathrm{L})}(\theta,M) \triangleq 1 - \frac{\theta \rho}{\rho_{0} R^{-\alpha}} \frac{(\pi \lambda)^{\frac{\alpha}{2}}}{\tfrac{\alpha}{2}-1} \frac{\big( M - \lceil \tfrac{\alpha}{2} \rceil \big)^{1 - \frac{\alpha}{2}}}{N_{\mathrm{R}} - M - 1}.
\end{align}
On the one hand, $\mathsf{P}_{\suc}^{(\mathrm{L})}(\theta,M)$ is concave in $M$ and its maximum is attained at
\begin{align}
\nonumber & \argmin_{\lceil \tfrac{\alpha}{2} \rceil + 1 \leq M \leq N_{\mathrm{R}} - 2} \frac{\big( M - \lceil \tfrac{\alpha}{2} \rceil \big)^{1 - \frac{\alpha}{2}}}{N_{\mathrm{R}} - M - 1} \\
& \hspace{2.5cm} = \bigg( 1 - \frac{2}{\alpha} \bigg) (N_{\mathrm{R}} - 1) + \frac{2}{\alpha} \ceil[\bigg]{\frac{\alpha}{2}}.
\end{align}
On the other hand, this lower bound -- based on Markov's inequality -- is quite loose (it even assumes negative values), especially for small values of $N_{\mathrm{R}}$ (as can be observed from Figures~\ref{fig:comp1}--\ref{fig:comp3}), and it is not defined for low values of $M$. For instance, if $\alpha = 4$ and $N_{\mathrm{R}} \leq 5$, $\mathsf{P}_{\suc}^{(\mathrm{L})}(\theta,M)$ cannot be computed for any $M$. As a matter of fact, these issues prevent us from using $\mathsf{P}_{\suc}^{(\mathrm{L})}(\theta,M)$ when analyzing small-cell networks, where the base stations are equipped with a low-to-moderate number of antennas. In the light of this, in the next section we propose two tight, tractable approximations of $\mathsf{P}_{\suc}(\theta,M)$.

\subsection{Tractable Approximations of the Success Probability} \label{sec:SP_appr}

The difficulty of obtaining a closed-form expression of the Laplace transform in \eqref{eq:LI} and the looseness of the lower bound in \eqref{eq:P_suc1_LB} prevent us from analyzing the success probability and optimizing the number of cancelled interfering nodes $M$. For this reason, let us introduce
\begin{align} \label{eq:d}
d_{M} \triangleq \Exp [X_{M}] = (\pi \lambda)^{-\frac{1}{2}} \frac{\Gamma \big( M + \tfrac{1}{2} \big)}{\Gamma(M)}
\end{align}
as the average distance between the reference receiver and the $M$-th nearest interfering node \cite{Hae05}. Then, we can approximate $I(M)$ in \eqref{eq:I} as
\begin{align}
\label{eq:I_t} \widetilde{I}(M) \triangleq \rho \sum_{i : X_{i} > d_{M}} X_{i}^{-\alpha} S_{i}(M)
\end{align}
and, consequently, $\SIR(M)$ in \eqref{eq:SIR} can be approximated as
\begin{align}
\label{eq:SIR_t} \widetilde{\SIR}(M) \triangleq \frac{\rho_{0} R_{0}^{-\alpha} S_{0}(M)}{\widetilde{I}(M)}.
\end{align}
Observe that, for a given realization of the PPP of the interfering nodes, exactly $M$ interfering nodes are cancelled in \eqref{eq:I}, whereas all the interfering nodes that fall within distance $d_{M}$ (which, on average, encloses $M$ nodes) are cancelled in \eqref{eq:I_t}. Therefore, $\widetilde{\mathsf{P}}_{\suc}(\theta,M) \triangleq \Pr[\widetilde{\SIR}(M) > \theta]$ approximates $\mathsf{P}_{\suc}(\theta,M)$ and has a tractable closed-form expression, as detailed in the following theorem. In Section~\ref{sec:num}, we show that this approximation is also very tight.

\begin{theorem} \label{th:P_suc2} \rm{
The success probability $\mathsf{P}_{\suc}(\theta,M)$ in \eqref{eq:P_suc1}, for $M \geq 1$, is approximated by
\begin{align} \label{eq:P_suc2}
\hspace{-1mm} \widetilde{\mathsf{P}}_{\suc}(\theta,M) = \! \! \sum_{n=0}^{N_{\mathrm{R}} - M - 1} \! \bigg[ \frac{(- s)^{n}}{n!} \frac{\mathrm{d}^{n}}{\mathrm{d} s^{n}} \setL_{\widetilde{I}(M)}(s) \bigg]_{s = \theta \rho_{0}^{-1} R_{0}^{\alpha}}
\end{align}
where
\begin{align} \label{eq:LI_t}
\setL_{\widetilde{I}(M)}(s) \triangleq \exp \big( - 2 \pi \lambda \Upsilon(s,M) \big)
\end{align}
is the Laplace transform of $\widetilde{I}(M)$ in \eqref{eq:I_t}, where we have defined
\begin{align} \label{eq:Upsilon}
\Upsilon(s, M) \triangleq \frac{s \rho d_{M}^{2-\alpha}}{\alpha-2} {}_2F_1 \big( 1, 1-\tfrac{2}{\alpha}, 2-\tfrac{2}{\alpha}, -s \rho d_{M}^{-\alpha} \big)
\end{align}
with ${}_2F_1(a,b,c,z)$ denoting the Gauss hypergeometric function \cite[Sec.~9.1]{Gra07}.}
\end{theorem}

\begin{IEEEproof}
See Appendix~\ref{sec:A_P_suc2_th}.
\end{IEEEproof} \vspace{1mm}

\noindent Note that the Laplace transform $\setL_{\widetilde{I}(M)}(s)$ in \eqref{eq:LI_t} is given in closed form; furthermore, its derivatives can be computed efficiently thanks to the property of the derivatives of the Gauss hypergeometric function by which $\frac{\diff^{n}}{\diff x^{n}} \,_{2}F_{1} \big(a,b,c,x) = \frac{(a)_{n} (b)_{n}}{(c)_{n}} \,_{2}F_{1} \big(a+n,b+n,c+n,x)$, with $(x)_{n} \triangleq \frac{\Gamma(x+n)}{\Gamma(x)}$. Although $\widetilde{\mathsf{P}}_{\suc}(\theta,M)$ can be easily evaluated numerically, its form is still complicated and it is difficult to observe the effect of the different parameters.

The following corollary gives a compact expression of the optimal success probability.

\begin{corollary} \label{cor:P_suc2_max} \rm{
The maximum of $\widetilde{\mathsf{P}}_{\suc}(\theta,M)$ is given by
\begin{align} \label{eq:P_suc2_max}
\max_{M} \widetilde{\mathsf{P}}_{\suc}(\theta,M) = \widetilde{\mathsf{P}}_{\suc}(\theta,M^{\star}) = \| \L(\theta) \|_{1}
\end{align}
where $M^{\star}$ denotes the optimal number of cancelled nodes, $\| \cdot \|_{1}$ is the $\ell_{1}$-induced matrix norm, and $\L(\theta) \in \Real^{N_{\mathrm{R}} \times N_{\mathrm{R}}}$ is a lower triangular matrix with elements $[\L(\theta)]_{i,j} \triangleq \Big[ \frac{(- s)^{j-1}}{(j-1)!} \frac{\mathrm{d}^{j-1}}{\mathrm{d} s^{j-1}} \setL_{\widetilde{I}(N_{\mathrm{R}}-i)}(s) \Big]_{s = \theta \rho_{0}^{-1} R_{0}^{\alpha}}$.}
\end{corollary}

The following upper bound on $\widetilde{\mathsf{P}}_{\suc}(\theta,M)$, obtained using Alzer's inequality \cite{Alz97}, provides an even more tractable approximation of $\mathsf{P}_{\suc}(\theta,M)$.\footnote{A lower bound on $\widetilde{\mathsf{P}}_{\suc}(\theta,M)$ can be obtained by fixing $\kappa_{M} = 1$ in \eqref{eq:P_suc2_UB}; however, such bound is not sufficiently tight and it is thus not considered.}

\begin{corollary} \label{cor:P_suc2_UB} \rm{
$\widetilde{\mathsf{P}}_{\suc}(\theta,M)$ can be upper bounded as $\widetilde{\mathsf{P}}_{\suc}(\theta,M) < \widetilde{\mathsf{P}}_{\suc}^{(\mathrm{U})}(\theta,M)$ with
\begin{align} \label{eq:P_suc2_UB}
\widetilde{\mathsf{P}}_{\suc}^{(\mathrm{U})}(\theta,M) \triangleq \sum_{n=1}^{N_{\mathrm{R}}-M} (-1)^{n-1} {{N_{\mathrm{R}}-M}\choose{n}} \setL_{\widetilde{I}(M)} (n \kappa_{M} s)
\end{align} \vspace{-1mm}

\noindent where we have defined $\kappa_{M} \triangleq \big(\Gamma(N_{\mathrm{R}}-M+1)\big)^{-\frac{1}{N_{\mathrm{R}}-M}}$ and with $s = \theta \rho_{0}^{-1} R_{0}^{\alpha}$.}
\end{corollary} \vspace{-1mm}

Having removed the derivatives, \eqref{eq:P_suc2_UB} provides more intuitive insights into the network performance with partial interference cancellation. However, the presence of the summation still prevents us from deriving a neat expression for the optimal number of cancelled nodes $M^{\star}$. Lastly, observe that, contrary to $\widetilde{\mathsf{P}}_{\suc}(\theta,M)$ in \eqref{eq:P_suc2}, $\widetilde{\mathsf{P}}_{\suc}(\theta,M)$ in \eqref{eq:P_suc2_UB} presents an alternating sum.


\section{Numerical Results} \label{sec:num}

In this section, we present numerical results to assess our theoretical findings. We begin by evaluating the obtained approximations of the success probability. The following parameters are used: the pathloss exponent is $\alpha = 4$, the transmit powers are $\rho_{0} = \rho = 1$~W, and the distance between the reference receiver and its associated transmitter is fixed to $R_{0} = 10$~m. The success probability $\mathsf{P}_{\suc}(\theta,M)$ (where exactly the nearest $M$ interfering nodes are cancelled) is obtained via Monte Carlo simulations with $10^{4}$ realizations of the PPP modeling the interfering nodes.

Figure~\ref{fig:comp1} plots $\widetilde{\mathsf{P}}_{\suc}(\theta,M)$ in \eqref{eq:P_suc2} and $\widetilde{\mathsf{P}}_{\suc}^{(\mathrm{U})}(\theta,M)$ in \eqref{eq:P_suc2_UB} against the number of cancelled nodes $M$ and with SIR threshold $\theta=1$. Two cases are considered: $\lambda = 10^{-2}$~nodes/m$^2$ with $N_{\mathrm{R}}=10$ and $\lambda = 2 \times 10^{-2}$~nodes/m$^2$ with $N_{\mathrm{R}}=20$. Contrary to the baseline lower bound $\mathsf{P}_{\suc}^{(\mathrm{L})}(\theta,M)$ in \eqref{eq:P_suc1_LB} (see \cite{Jin08}), the proposed $\widetilde{\mathsf{P}}_{\suc}(\theta,M)$ tightly approximates the success probability $\mathsf{P}_{\suc}(\theta,M)$; furthermore, its upper bound $\widetilde{\mathsf{P}}_{\suc}^{(\mathrm{U})}(\theta,M)$ is also tight. In addition, in these settings, imposing $M = 0$ (no cancellation) or $M = N_{\mathrm{R}}-1$ (full zero forcing) yields approximately the same network performance.

\begin{figure}[t!]
\centering
\includegraphics[scale=1]{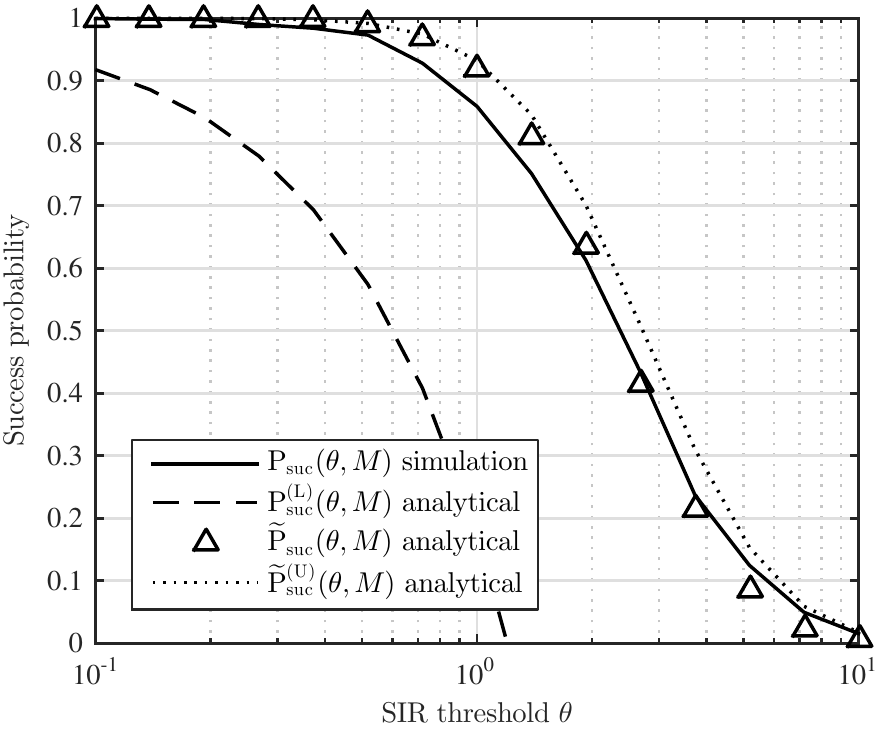} \vspace{-2mm}
\caption{Success probability versus SIR threshold $\theta$ with $N_{\mathrm{R}}=10$, $M=5$, and $\lambda = 10^{-2}$~nodes/m$^2$.} \label{fig:comp3} \vspace{4mm}
\end{figure}

Figures~\ref{fig:comp2} and \ref{fig:comp3} consider $N_{\mathrm{R}}=10$ and $M=5$ and plot the success probability and their approximations against the density $\lambda$ and the SIR threshold $\theta$, respectively. On the one hand, we note that $\mathsf{P}_{\suc}^{(\mathrm{L})}(\theta,M)$ in \eqref{eq:P_suc1_LB} is decreasingly tight as $\lambda$ and $\theta$ increase: therefore, $\mathsf{P}_{\suc}^{(\mathrm{L})}(\theta,M)$ is particularly unsuitable for studying UDNs. Moreover, we observe that $\widetilde{\mathsf{P}}_{\suc}(\theta,M)$ upper bounds $\mathsf{P}_{\suc}(\theta,M)$ for low values of $\lambda$ and $\theta$, while it lower bounds $\mathsf{P}_{\suc}(\theta,M)$ for high values of $\lambda$ and $\theta$.

\begin{figure}[t]
\centering
\includegraphics[scale=1]{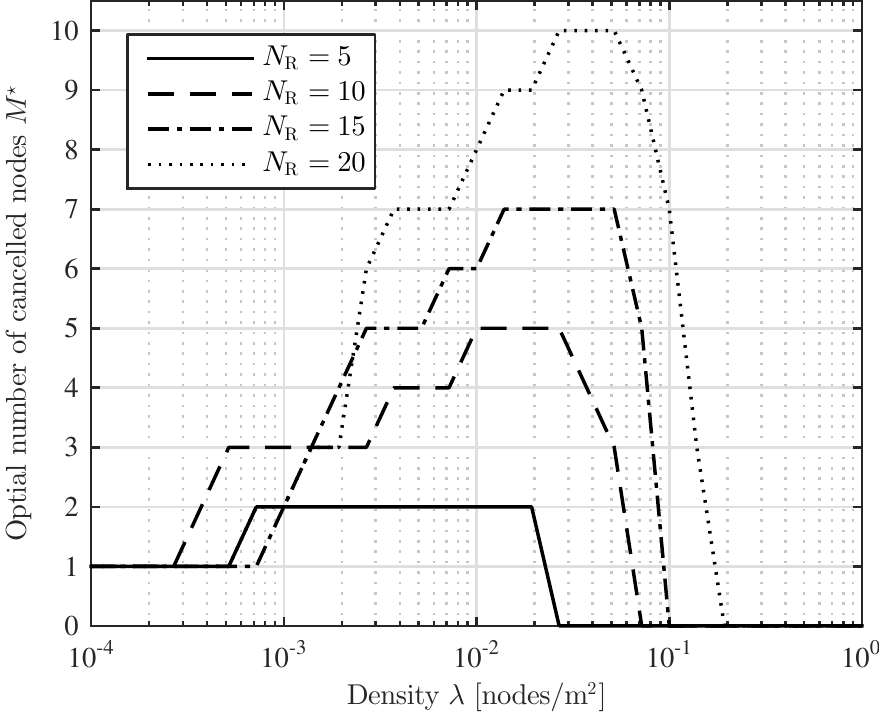} \vspace{-2mm}
\caption{Optimal number of cancelled nodes $M^{\star}$ versus density $\lambda$ with $\theta = 1$.} \label{fig:M_opt}
\end{figure}

Lastly, Figure~\ref{fig:M_opt} shows the optimal number of cancelled nodes $M^{\star}$ against the density $\lambda$ for different values of $N_{\mathrm{R}}$. Interestingly, we observe that $M^{\star} \leq \lfloor \frac{N_{\mathrm{R}}}{2} \rfloor$, which means that no more than half of the available degrees of freedom should be used for interference cancellation. Furthermore, as expected, we have that $M^{\star} = 0$ (i.e., MRC becomes optimal) at high densities.

\section{Conclusions} \label{sec:concl}


In this paper, we presented preliminary results on the performance of ultra-dense networks with multiple antennas and partial interference cancellation at the receiver. More specifically, we provided both integral expressions and tight tractable approximations for the probability of successful transmission, which can be used to accurately evaluate the optimal tradeoff between interference cancellation and array gain. Numerical results showed that no more than half of the available degrees of freedom should be used for interference cancellation.

\appendices

\section{Proof of Theorem~\ref{th:P_suc1}} \label{sec:A_P_suc1_th}

Given the interference term $I(M)$ defined in \eqref{eq:I}, the success probability is given by
\begin{align}
\mathsf{P}_{\mathrm{suc}}(M) & = \Pr \bigg[ \frac{\rho_{0} R_{0}^{-\alpha} S_{0}(M)}{I(M)} > \theta \bigg] \\
& = \Pr \big[ S_{0}(M) > \theta \rho_{0}^{-1} R_{0}^{\alpha} I(M) \big] \\
& = \Exp_{I(M)} \big[ \bar{F}_{S_{0}(M)} \big( \theta \rho_{0}^{-1} R_{0}^{\alpha} I(M) \big) \big].
\end{align}
Since the reference receiver is equipped with $N_{\mathrm{R}}$ antennas and applies PZF to cancel the nearest $M$ interfering nodes, we have that: i) the signal from its associated transmitter is distributed as $S_{0}(M) \sim \chi_{2 (N_{\mathrm{R}}-M)}^{2}$;\footnote{We define a random variable $X \sim \chi_{2 N}^{2}$ to have PDF $f_{X}(x) = \frac{x^{N-1} e^{-x}}{\Gamma(N)}$; its CCDF is given by $\bar{F}_{X}(x) = 1- \frac{\gamma(N,x)}{\Gamma(N)} = e^{-x} \sum_{n=0}^{N-1} \frac{x^{n}}{n!}$.} ii) the interference from the uncancelled interfering nodes is distributed as $S_{i}(M) = S_{i} \sim \chi_{2}^{2}$, $i=M+1, \ldots, \infty$. Therefore, our case falls into the general framework \cite{Hun08} and $\mathsf{P}_{\suc}(\theta,M)$ in \eqref{eq:P_suc1} results from applying \cite[Th.~1]{Hun08} (see footnote 3)
\begin{align}
\nonumber & \hspace{-1.1cm} \Exp_{I(M)} \big[ \bar{F}_{S_{0}(M)} \big( s I(M) \big) \big] \\
& = \Exp_{I(M)} \bigg[ e^{- s I(M)} \sum_{n=0}^{N_{\mathrm{R}}-M-1} \frac{\big( s I(M) \big)^{n}}{n!} \bigg] \\
& = \sum_{n=0}^{N_{\mathrm{R}}-M-1} \bigg[ \frac{(-s)^{n}}{n!} \frac{\diff^{n}}{\diff s^{n}} \setL_{I(M)}(s) \bigg]
\end{align}
where the Laplace transform of $I(M)$ is given by
\begin{align}
\hspace{-2mm} \setL_{I(M)}(s) & = \Exp \big[ e^{-s I(M)} \big] \\
& = \Exp \Bigg[ \exp \bigg( - s \rho \sum_{i = M+1}^{\infty} X_{i}^{-\alpha} S_{i}(M) \bigg) \Bigg] \\
& = \Exp_{\Phi} \Bigg[ \prod_{i = M+1}^{\infty} \! \Exp_{S_{i}} \big [ \exp \big( - s \rho X_{i}^{-\alpha} S_{i}(M) \big) \big] \Bigg].
\end{align}
Then, the expression in \eqref{eq:LI} is obtained by applying the MGF of the exponential distribution. In the case of $M=0$, we can further apply the PGFL of a PPP, which gives
\begin{align}
\hspace{-1mm} \setL_{I(0)}(s) = \exp \bigg( -2 \pi \lambda \int_{0}^{\infty} \bigg( 1- \frac{1}{1+s \rho r^{-\alpha}} \bigg) r \diff r \bigg).
\end{align}
Solving the above integral, we obtain the expression in \eqref{eq:LI_0}. This completes the proof. \hfill \IEEEQED

\section{Proof of Theorem~\ref{th:P_suc2}} \label{sec:A_P_suc2_th}

Given the interference term $\widetilde{I}(M)$ defined in \eqref{eq:I_t}, the success probability is given by
\begin{align}
\widetilde{\mathsf{P}}_{\mathrm{suc}}(M) & = \Pr \bigg[ \frac{\rho_{0} R_{0}^{-\alpha} S_{0}(M)}{\widetilde{I}(M)} > \theta \bigg].
\end{align}
Therefore, the expression of $\widetilde{\mathsf{P}}_{\suc}(\theta,M)$ in \eqref{eq:P_suc2} is obtained by following similar steps as in the proof of Theorem~\ref{th:P_suc1}. On the other hand, the Laplace transform of $\widetilde{I}(M)$ is given by
\begin{align}
\hspace{-2mm} \setL_{\widetilde{I}(M)}(s) & = \Exp \big[ e^{-s \widetilde{I}(M)} \big] \\
& = \Exp \Bigg[ \exp \bigg( - s \rho \sum_{i : X_{i} > d_{M}} X_{i}^{-\alpha} S_{i}(M) \bigg) \Bigg] \\
& = \Exp_{\Phi} \Bigg[ \prod_{i : X_{i} > d_{M}} \! \! \Exp_{S_{i}} \big [ \exp \big( \! - s \rho X_{i}^{-\alpha} S_{i}(M) \big) \big] \Bigg] \\
& = \Exp_{\Phi} \Bigg[ \prod_{i : X_{i} > d_{M}} \frac{1}{1+s \rho X_{i}^{-\alpha}} \Bigg] \\
& = \exp \bigg( -2 \pi \lambda \int_{d_{M}}^{\infty} \bigg( 1- \frac{1}{1+s \rho r^{-\alpha}} \bigg) r \diff r \bigg).
\end{align}
Finally, solving the above integral, we obtain the expression in \eqref{eq:LI_t}. This completes the proof. \hfill \IEEEQED

\vspace{2mm}

\addcontentsline{toc}{chapter}{References}
\bibliographystyle{IEEEtran}
\bibliography{IEEEabrv,ref_Huawei}

\end{document}